\pdfoutput=1
\documentclass{article}
\usepackage{spconf,amsmath,graphicx}

\title{TT-Net: Dual-path transformer based sound field translation in the spherical harmonic domain}
\name{Yiwen Wang, Zijian Lan, Xihong Wu, Tianshu Qu}
\address{Key Laboratory on Machine Perception (Ministry of Education) \\School of Intelligence Science and Technology \\Peking University, Beijing, China\\
\{pku\_wyw, qutianshu\}@pku.edu.cn}
\begin{document}
\ninept
\bibliographystyle{ieeetr} 
\maketitle
\begin{abstract}
In the current method for the sound field translation tasks based on spherical harmonic (SH) analysis, the solution based on the additive theorem usually faces the problem of singular values caused by large matrix condition numbers. The influence of different distances and frequencies of the spherical radial function on the stability of the translation matrix will affect the accuracy of the SH coefficients at the selected point. Due to the problems mentioned above, we propose a neural network scheme based on the dual-path transformer. More specifically, the dual-path network is constructed by the self-attention module along the two dimensions of the frequency and order axes. The transform-average-concatenate layer and upscaling layer are introduced in the network, which provides solutions for multiple sampling points and upscaling. Numerical simulation results indicate that both the working frequency range and the distance range of the translation are extended. More accurate higher-order SH coefficients are obtained with the proposed dual-path network.
\end{abstract}
\section{Introduction}
\label{sec:introduct}
With sound field recording, processing, and reproduction techniques, spatial audio enables the reconstruction of an acoustic environment \cite{cobos2022overview}. The high-order ambisonics (HOA) can realize the expression of the three-dimensional sound field within a specific range \cite{ poletti2005three}. The higher-order coefficients can be acquired through spherical microphone array (SMA) recordings \cite{abhayapala2002theory}. Virtual reality applications demand additional translation degrees of freedom, often referred to as \textit{six degrees of freedom} (6DoF). 

A common way to implement 6DoF is to use multiple SMAs distributed in three-dimensional space. The commonly used method uses the mathematical properties of SH to convert the SH coefficients obtained at the sampling points into the higher-order SH coefficients at the selected point. In \cite{laborie2003new}, Laborie \textit{et al.} present the theoretical principle for estimating the SH coefficients, including the spatial sampling and encoding aspects. Based on the translation theorem, Samarasinghe \textit{et al.} provide the solution for the global coefficients using higher-order microphones (HOM) \cite{samarasinghe2014wavefield}. In \cite{rafaely2015fundamentals, peleg2011investigation}, according to the category and location information of the sound source, Rafaely \textit{et al.} summarize the translation equation corresponding to different situations. However, the numerical method is relatively complex and suffers from the problem of ill-conditioned matrices under higher-order conditions. Based on the idea of plane wave decomposition, Wang \textit{et al.} propose a solution without using the translation theorem \cite{wang2018translations}. Ueno \textit{et al.} formulate an estimate of the harmonic coefficients based on infinite-order analysis by applying Bayes’ theorem \cite{ueno2017sound, nakanishi2019two}. Both methods have achieved better results in cylindrical coordinates, but the effect needs to be improved as the frequency increases. Furthermore, there is a lack of experimental validation for the case in the 3D space.

The ability of deep learning to model complex relationships between different representations has been applied to SH-based sound field problems recently. To achieve higher bandwidth SH coefficients and alleviate spatial aliasing problems, networks are used to model SH bases \cite{gao2022sparse}. In the previous work, a U-Net-based generator is used to realize the upscaling of SH coefficients \cite{wang2022up}. Given the limitations of the traditional numerical solution, a neural-network-based SH coefficient translation is proposed to achieve a more accurate SH representation at different distances, different frequency bands, and under complex sound source conditions in our work. The experimental results show that the proposed network-based method extends the working frequency range and obtains more precise translation coefficients under complex environmental conditions.

The rest of the paper is organized as follows: Section 2 introduces the theory of translation matrix in the SH domain, and Section 3 describes the proposed dual-path transformer model. Experimental setup and results are reported in Section 4 and Section 5. Finally, we conclude in Section 6.

\section{Theory of translation matrix}
\label{sec:translatetheory}


Consider the external far-field sound source situations. The sound pressure in the spherical coordinate system can be decomposed into the expansion of SH coefficients as 
\begin{equation}
p(k, \textbf{r}) = \sum_{n=0}^{\infty} j_{n}(kr) \sum_{m=-n} ^{n} B_{n}^{m} Y_{n}^{m}(\theta, \phi), 
\label{equ:sound_pressure_expression}
\end{equation}
where $k$ is the wave number, $\textbf{r} \equiv (r, \theta, \phi)$ is the  spherical coordinate system denoted by elevation and azimuth angles, $\theta$ and $\phi$, together with the radial distance $r$, $p(k, \textbf{r})$ represents the sound pressure at $\textbf{r}$, $j_{n}(kr)$ is the spherical Bessel function, $Y_n^m(\theta,\phi)$ is the basis function of SH, and $B_{n}^{m}$ is the corresponding SH coefficient. Consider the translation from a global origin to a local translated origin $\textbf{r}^{''} \equiv (r^{''}, \theta^{''}, \phi^{''})$, 
such that 
\begin{equation}
    \textbf{r} = \textbf{r}^{''} + \textbf{r}^{'},
    \label{equ:translation_coeff}
\end{equation}
where $\textbf{r}$ and $\textbf{r}^{'} \equiv (r^{'}, \theta^{'}, \phi^{'})$ represent the position relative to the original global origin and the new coordinate center, respectively. 


According to the addition theorems \cite{chew1995waves}, the translation from the spherical Bessel functions to spherical Bessel functions is described as 
\begin{equation}
\begin{aligned}
    j_{n}(k r)Y_{n}^{m} = & \sum_{n^{'}=0}^{\infty} \sum_{m^{'} = -n^{'}}^{n^{'}}j_{n^{'}}(k r^{'}) j_{n^{'}}(k r^{'}) Y_{n^{'}}^{m^{'}}(\theta^{'}, \phi^{'}) \\ 
    & \times \sum_{n^{''} = 0}^{\infty}j_{n^{''}}(k r^{''})Y_{n^{''}}^{m-m^{'}}(\theta^{''}, \phi^{''})C_{n^{'} m^{'}} ^{n m n^{''}},
    \label{equ:addition_theory}
\end{aligned}
\end{equation}
where $C_{n^{'} m^{'}} ^{n m n^{''}}$  contains the multiplication of two Wigner 3-j operators. See \cite{rafaely2015fundamentals} for details.

For each translation position $\textbf{r}^{''}$, the sound pressure is expressed by a set of SH coefficients with $\textbf{r}^{''}$ as the local coordinate center. Assuming that the SH coefficients of the $Q$ local coordinate systems are truncated to order $N^{''}$, and the coefficients of the global origin are truncated to order $N$. According to the addition theorems, the relationship between the two types of coefficients can be established through the translation matrix of the SH coefficients. The derivation of this part is described in detail in \cite{rafaely2015fundamentals}. The formula is expressed as 
\begin{equation}
    (b^{''})^{T} = T_{trans} b^{T},
\end{equation}
where $b^{''} = [B_{0(0)}^{0}, B_{1(0)}^{-1}, B_{1(0)}^{{0}}, ...,B_{n(q)}^{m},... B_{N^{''}(Q-1)}^{N^{''}}]$ is a $1 \times (Q(N^{''}+1)^{2})$ vector, $B_{n(q)}^{m}$ represents the local spherical coefficient of the order $n$ and degree $m$ for the $q$-th local coordinate system, $b = [B_{0}^{0}, B_{1}^{-1}, B_{1}^{{0}}, ..., B_{n}^{m}...,B_{N}^{N}]$ is a $1\times (N+1)^2$ vector, $B_{n}^{m}$ represents the global coefficient. The shape of the translation matrix $T_{trans}$ is $Q(N^{''}+1)^{2} \times (N+1)^2$. Each row element of the matrix is expanded according to the order of the global origin system.

\section{Proposed method}
\label{sec:method}

In the traditional method, the relationship of SH coefficients of the different coordinate centers is established through the translation matrix. Theoretically, the coefficients of the global origin can be realized by matrix inversion operation. However, in practice, due to the problem of Bessel nulls \cite{samarasinghe2014wavefield}, the matrix has a large condition number, which seriously affects the results of numerical calculation. Ridge regression can alleviate the problem of matrix singularity to a certain extent \cite{mcdonald2009ridge}, but in practical applications, different distance conditions bring errors to the coefficients at different orders. We propose a dual-path transformer-based SH coefficient translation network to solve the above problems in our work. 

\subsection{Model Description}

\begin{figure}[htbp]
    \centering
    \includegraphics[height=28.7mm]{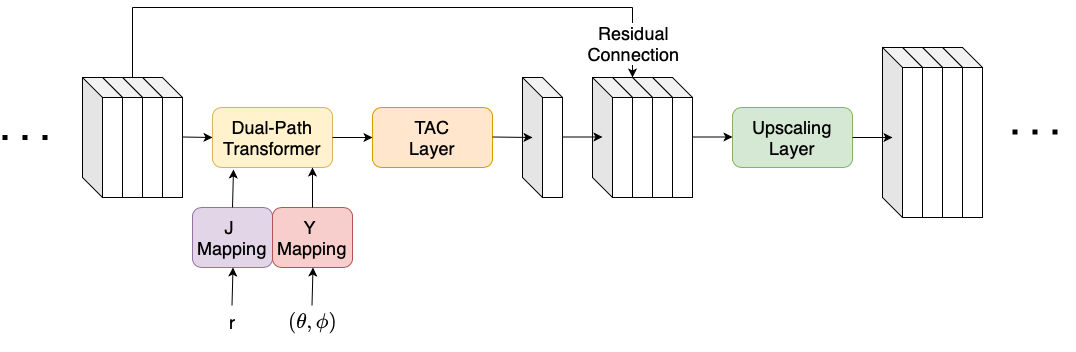}
    \caption{Single-layer architecture of TT-Net.}
    \label{fig:TT_model}
\end{figure}

A dual-path transformer-based network for translation named TT-Net is proposed.  More specifically, we take a sample at one location as an example to illustrate our proposed network architecture in detail. One layer of the TT-Net is shown in Fig. \ref{fig:TT_model}. The input and output of the network are SH coefficients of different orders. In our work, the input feature of SH coefficients is a $K \times (N+1)^{2}$ matrix, where $K$ is the total number of the frequency bins, and $N$ is the order of the SH. The transformer mentioned in the paper refers to the encoder part. As mentioned in Eq. (\ref{equ:addition_theory}), for the SH translation, both radial spherical Bessel functions and angle-dependent SH functions contribute to the translation process.  Due to the decoupling of distance and angle, two mapping networks named $J$ and $Y$ are constructed to replace the expression of $J_{n} (kr)$ and $Y_{n}^{m} (\theta, \phi)$, respectively. $Y_{n}^{m} (\theta, \phi)$ is only related to order and orientation, not frequency. Therefore, the same constraints are used for the $Y$ network, the input is the angle information, and the output is the $N$-order vector after the reshaping operation. Similar structural constraints are applied to the $J$ network so that the output dimension of the $J$ network is consistent with the dimension of the spherical Bessel function, that is, $k\times (N+1)$.  The output of the Dual-Path Transformer (DPT) remains the same shape as the input SH coefficients. The output of the DPT module serves as the input to a transform-average-concatenate (TAC) layer \cite{luo2020end} and a fully connected (FC) layer. The TAC layer is used to integrate different SH coefficients, while the FC layer is used for upscaling the coefficients \cite{routray2019deep}. A residual connection is added between the TAC layer and the Upscaling Layer to help the training of the network \cite{he2016deep}. 

The network is connected by the structure shown in Fig. \ref{fig:TT_model}. The order of the output of every single layer is larger than the input.  The last layer of the network no longer uses the residual connection and upscaling, and the output of TAC from the last layer is used as the final output of the entire model.

\subsection{Dual-path Transformer Module}

\begin{figure}[htbp]
    \centering
    \includegraphics[height=29.2mm]{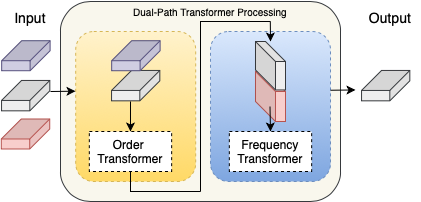}
    \caption{Dual-path Transformer Module.}
    \label{fig:dual_path_transformer}
\end{figure}

Dual-path transformer module has recently attracted much attention and has achieved good performance in speech separation \cite{chen2020dual, wang2021tstnn} and noise reduction \cite{dang2022dpt}. The dual-path module used in our work is shown in Fig. \ref{fig:dual_path_transformer}. The input conditions represented by different colors are consistent with Fig. \ref{fig:TT_model}. Taking the spherical Bessel function as an example, for the same frequency bin $k$, $J_{n}(kr)$ of different orders is related to the SH coefficients of the same frequency bin. The coupling of coefficients and radial spherical functions at the same frequency bin is achieved through the self-attention mechanism. Compared with the standard encoder part in the transformer, a fully-connected layer is added to the last layer of the encoder to integrate the SH coefficient and the $J$ function. It also plays a role in constraining the same input and output dimensions. The output of the intermediate layer is transposed and concatenated with the $Y$ function, and a similar self-attention operation is performed, followed by the FC layer.


\section{EXPERIMENTAL SETUP}
\subsection{Datasets And Training Settings}
During the translation process, information such as location, frequency, and the number of sampling points will affect the results. Therefore, different situations are taken into account in the simulation data. For our subsequent applications, we use the fourth-order SH coefficients as input, which means $N=4$. The frequency bin ranges from 100Hz to 3000Hz with an interval of 100Hz, which means $K$=30. The coefficient of the global origin is $N^{''} = 8$. The distance between the set sampling point and the global origin is randomly sampled between 0.2$m$ and 2.0$m$. In our experiment, plane waves from 1 to 4 directions are randomly generated as the signal source. The amplitude of the signal is randomly chosen from 0.1 to 1.0. Since the low-frequency system noise brings difficulties to the solution of SH coefficients \cite{lin2020anti}, noises of different signal-to-noise ratios (SNR) are added to enhance the noise immunity of the model. SNR varies from 10 to 30dB. The number of spatial sampling points is set from 4 to 10. The minimum setting is four sampling points to ensure the full rank of the translation matrix. 

Networks are trained with mean squared error (MSE) loss. The order of the output of each dual-path transformer module is one greater than the input. Multi-head attention is introduced, and the number of heads is the same as the current order plus one. For stable learning, gradients are clipped to [-1.0, 1.0]. All models are trained distributedly with 2 TITAN RTX with batch size 32.  The learning rate is initialized to be 3e-4 and halved with Adam optimizer training \cite{kingma2014adam}. For training stability, each data is normalized, and the number of data sampling points in each batch is ensured to be the same. The training data are trained sequentially from 10 to 4.

\subsection{Evaluation Metrics}
The evaluation is based on the similarity of the SH coefficients and the sound field. Euclidean distance metric (EDM) and cosine similarity (COSS) metrics are used to judge the difference between the recovered and the ideal coefficients, respectively, where EDM gives judgments from Euclidean space, and COSS provides judgments of structural similarity.
\begin{equation}
    EDM = \frac{1}{N} \sum_{i=1}^{N} |\hat{y}_{i} - y_{i}|^{2},
\end{equation}
\begin{equation}
    COSS = \frac{1}{N} \sum_{i=1}^N \frac{\hat{y}_{i} \cdot y_{i}}{ |\hat{y}_{i}| \times |y_{i}|},
\end{equation}
$N$ is the total number of test data, $y$ is the SH coefficients of the global origin, and $\hat{y}$ is the estimated result. Besides, signal-to-distortion ratio (SDR) is for evaluating the sound field based on the SH coefficients. SDR is defined as \cite{ueno2017sound}, 
\begin{equation}
    SDR = \frac{1}{N} \sum_{i=1}^N 10 log_{10} \frac{\int_{r\in V} u_{i}(\mathbf{r})^2 d \mathbf{r}}{\int_{r\in V} |u_{i}(\mathbf{r})-\hat{u}_{i}(\mathbf{r})|^2 d \mathbf{r}},
\end{equation}
where $N$ is the total number of test data, $u$ and $\hat{u}$ represent the sound pressure reconstructed with the ideal SH coefficients and the estimated coefficients. $SDR$ is calculated within a radius of 1.00$m$ at 0.02$m$ intervals. $k$ representing frequency is omitted.

\section{EVALUATION RESULTS AND DISCUSSION}
\label{sec:experiment}

\subsection{Ablation Study}

\begin{table}[htbp]
\caption{Average results of 8 spatial sampling points.}
\begin{tabular}{c|c|ccc}
\hline
Method & Strategy       & COSS           & EDM            & SDR(dB)        \\ \hline
LSM   & Regularization & 0.101          & 0.068          & -0.074         \\ \hline
TT-Net(1)  & Lrg2Sml L2     & 0.334          & 0.058          & 0.827          \\
TT-Net(2)  & Lrg2Sml L2     & 0.528          & 0.049          & 1.357          \\
\textbf{TT-Net(4)} & \textbf{Lrg2Sml  L2}    & \textbf{0.732} & \textbf{0.043} & \textbf{2.037} \\
TT-Net(4) & Lrg2Sml L1     & 0.472          & 0.052          & 1.397          \\
TT-Net(4) & Sml2Lrg L2     & 0.411          & 0.056          & 0.793          \\ 
 \hline
\end{tabular}
\label{tb:ablation_study}
\end{table}

The results of the ablation study are shown in Table \ref{tb:ablation_study}. The test data uses eight randomly selected spatial sampling points in the 3D space with a $1m$ distance. The direction of the single sound source is randomly generated in the 3D space. The number of test datasets is 600. The result is the average metrics of all frequencies. The method of solving by inversion of the translation matrix in \cite{peleg2011investigation} is abbreviated as the least square method (LSM). One-layer and two-layer architectures shown in Fig. \ref{fig:TT_model} are used for comparison to verify the optimal effect of increasing the dual-path module step by step. Correspondingly, the upscaling parts implement the mapping from order 4 to 8 or from other 4 to 6 with eight behind. The numbers in parentheses are the layers of the architecture. The number 4 indicates the scheme is that the order increases sequentially from 4 to 8. Lrg2Sml means that the network is trained with the number of spatial sampling points from 10 to 4, while Sml2Lrg is the opposite. L1 loss is chosen for comparison. The results show that Lrg2Sml helps the training. Results also show that the performance decreases if fewer DPT module is used.  Although L1 loss function performs better in some regression tasks \cite{ren2015faster}, this is not the case in our work. The final average results show that our proposed method is effective regarding SH coefficients and the recovered sound field.

\begin{figure*}[htb]
\begin{minipage}[b]{.46\linewidth}
  \centering
  \centerline{\includegraphics[width=9.45cm]{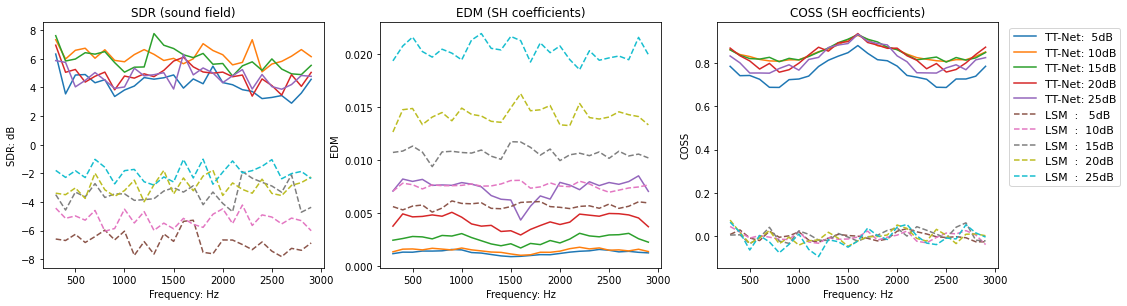}}
  \centerline{(a) Comparison for different SNRs.}\medskip
\end{minipage}
\hfill
\begin{minipage}[b]{0.46\linewidth}
  \centering
  \centerline{\includegraphics[width=9.45cm]{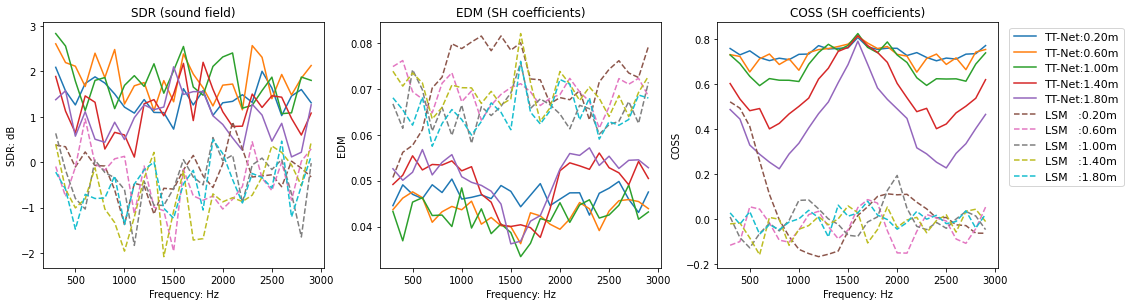}}
  \centerline{(b) Comparison for different distances.}\medskip
\end{minipage}
%
\begin{minipage}[b]{.46\linewidth}
  \centering
  \centerline{\includegraphics[width=9.45cm]{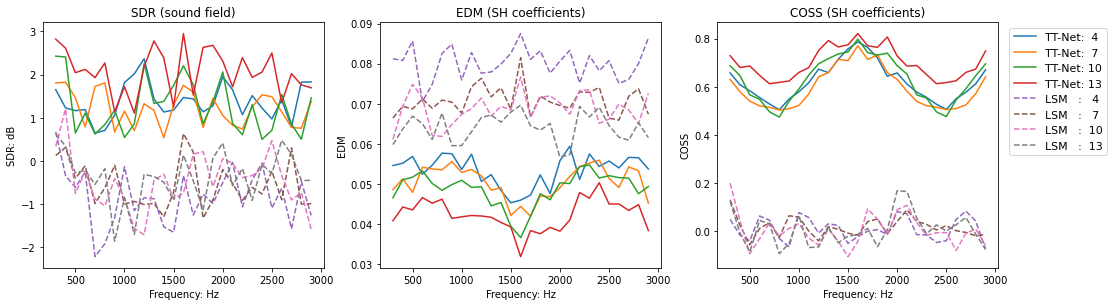}}
  \centerline{(c) Comparison for different sampling nums.}\medskip
\end{minipage}
\hfill
\begin{minipage}[b]{0.46\linewidth}
  \centering
  \centerline{\includegraphics[width=9.45cm]{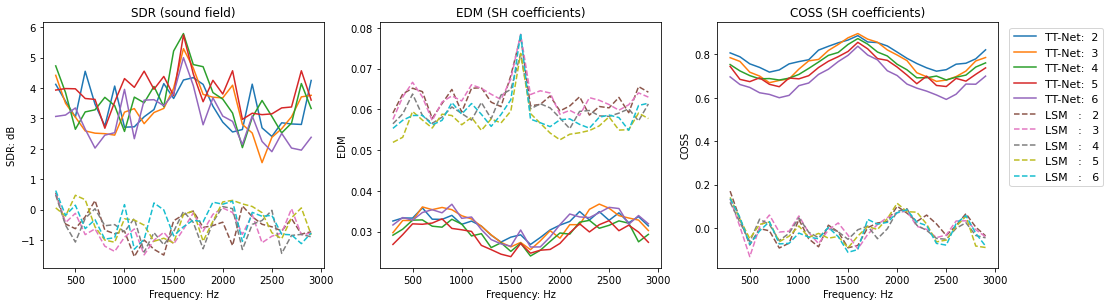}}
  \centerline{(d) Comparison for multiple sound sources.}\medskip
\end{minipage}
\caption{Experiments results under different conditions.}
\label{fig:Results}
\end{figure*}



\subsection{Experiments Results}
\subsubsection{Results for different SNRs}

The results for different SNRs are shown in Fig. \ref{fig:Results}(a).  From left to right are the results of SDR, EDM, and COSS. Horizontal is the frequency information. The TT-Net(4) with the best results is used, and we use TT-Net for short. LSM is compared with our optimal scheme. From the SDR results, it can be seen that the performance of the LSM is degraded as the SNR decreases. In contrast, our method performs consistently on SDR. The metrics of EDM and COSS show that the results of the coefficients are in line with the previous discussion. The noises cause outliers to be added to the numerical solution results, which seriously degrades the results. On the contrary, our proposed method results in noise immunity. 

\subsubsection{Results for different distances}
The results for different distances of spatial sampling points are shown in Fig. \ref{fig:Results}(b). Each test data uses eight randomly distributed sampling points on an equidistant sphere. The number of test samples per distance is 100. COSS shows that the traditional method works well at low frequencies. As the frequency increases, the results turn worse due to the increase of $kr$, which is related to the properties of spherical Bessel functions. The network method has stable performance under a shorter than 1.00$m$ condition. However, the performance degrades under long-distance conditions, and there exists singularity in the frequency band below 1kHz and above 2kHz. According to the analysis of the SDR results, the reconstructed sound field results remain the same, which indicates that the results of singular values are located in the higher-order part leading to little effect on the sound field near the origin.


\subsubsection{Results for different nums of spatial sampling points}

For experiments with different spatial sampling points, 4 to 13 spatial sampling points are  selected, and the number of test samples for each condition is 100. We select four cases, with the number of 4, 7, 10, and 13 spatial sampling points for visualization. In LSM, using more spatial sampling points reduces the singularity of the translation matrix, which is reflected in all three metrics, as shown in Fig. \ref{fig:Results}(c). As the number increases, the network also has a consistent trend. Under the test conditions, the number outside the training set achieves better results. The results show that the proposed method can increase the stability as the number increases and confirms the TAC module's role. 

\subsubsection{Results for multiple sound sources}

The results for the multi-sources condition are shown in Fig. \ref{fig:Results}(d). In LSM, the influence of sound sources in different directions only affects the SH coefficients, which does not affect the solution of the translation matrix. Therefore, the traditional method is consistent. That is, the number of sound sources has little effect. The network is tested with 2 to 6 different numbers of sound sources. The results show that the SH coefficients of different numbers of sound sources tend to be consistent. The performance of our method does not degrade as the number of sound sources increases. The test sets with more than four sound sources are consistent with the other results on the metrics of EDM and COSS. It should be noted that there are specific differences in the results of the proposed method at different frequencies, which will be further analyzed in the follow-up work.

\begin{figure}[htbp]
\begin{minipage}[b]{.99\linewidth}
  \centering
  \centerline{\includegraphics[height=2.53cm]{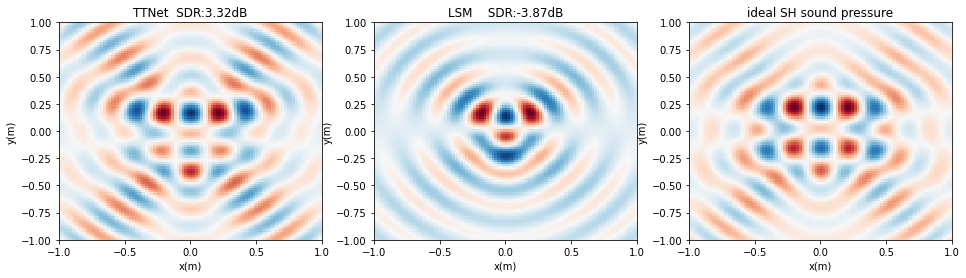}}
  \centerline{(a) 1000Hz.}\medskip
\end{minipage}
\hfill
\begin{minipage}[b]{.99\linewidth}
  \centering
  \centerline{\includegraphics[height=2.53cm]{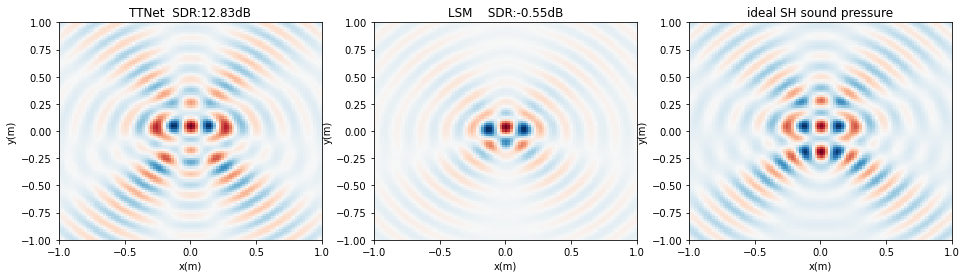}}
  \centerline{(b) 1800Hz.}\medskip
\end{minipage}
\caption{Visualization of the sound pressure in the x-y plane.}
\label{fig:visulize_sp}
\end{figure}

Fig. \ref{fig:visulize_sp} visualizes an example of the sound pressure at 1000 Hz and 1800 Hz. The figure shows the sound pressure distribution on a horizontal plane of $2m \times 2m$. Two sound sources in this example are oriented at 0$^\circ$ and 225$^\circ$. Sound pressures expanded by the output of TT-Net, LSM, and ideal SH coefficients are shown from left to right in each case. The results show that the network's performance is stable and better under different frequency conditions.

\section{Conclusion}
\label{sec:conclusion}
We propose a sound field recording method based on a dual-path transformer network. The method applies to the translation of SH coefficients. This work continues our exploration of optimizing SH analysis using neural networks. The proposed method reduces the occurrence of singular solutions in solving SH coefficients. The simulation results of different frequency ranges, SNRs, and under more complex sound source conditions show that the method is more accurate than the traditional method to recover the SH coefficients. Under the same conditions, the proposed method brings a 3dB improvement in the SDR metrics. Detailed studies remain as future work for applications in real-world scenarios containing multiple scattering rigid balls. Future work will be tested in real scenarios to provide a feasible solution for the realization of 6DoF.

\vfill\pagebreak




\bibliographystyle{IEEEbib}
\bibliography{reference}

\end{document}